\documentclass[twocolumn,showpacs,prl]{revtex4-2}

\usepackage{textcomp}
\usepackage{amsmath}
\usepackage{amssymb}
\usepackage{graphicx}
\usepackage{subfigure}
\usepackage{xcolor}

\makeatletter

\providecommand{\tabularnewline}{\\}
\makeatother

\begin{document}

\title{Design principles of nonlinear optical materials for Terahertz lasers}

\author{Juan Han$^{1,3}$}
\author{Yiwei Sun$^{1,3}$}
\author{Xiamin Huang$^{1,3}$}
\author{Wenjun Shuai$^{1,3}$}
\author{Guangyou Fang$^{1,2,3}$}
\author{Zhou Li$^{1,2,3,4}$}
\email{zli5@ualberta.ca}

\affiliation{$^{1}$ GBA Branch of Aerospace Information Research Institute, Chinese Academy of Sciences, Guangzhou, 510700, China \\
$^{2}$ University of Chinese Academy of Sciences, Beijing 100039, China \\
$^{3}$ Guangdong Provincial Key Laboratory of Terahertz Quantum Electromagnetics, Guangzhou 510700, China \\
$^{4}$ National Institute for Materials Science (NIMS), Tsukuba, Ibaraki 305-0044, Japan}
\begin{abstract}
We have investigated both inter-band and intra-band second order nonlinear optical conductivity based on the velocity correlation formalism and the spectral expansion technique. We propose a scenario in which the second order intra-band process is nonzero while the inter-band process is zero. This occurs for a band structure with momentum asymmetry in the Brillouin zone. Very low-energy photons are blocked by the Pauli exclusion principle from participating in the inter-band process; however, they are permitted to participate in the intra-band process, with the band smeared by some impurity scattering. We establish a connection between the inter-band nonlinear optical conductivity in the velocity gauge and the shift vector in the length gauge for a two-band model. Using a quasiclassical kinetic approach, we demonstrate the importance of intra-band transitions in high harmonic generations for the single tilted Dirac cone model and hexagonal warping model. We confirm that the Kramers-Kronig relations break down for the limit case of ($\omega$, $-\omega$) in the nonlinear optical conductivity. Finally, we calculate the superconducting transition temperature of NbN and the dielectric function of AlN, and the resistance of the NbN/AlN junction. The natural non-linearity of the Josephson junction brings a Josephson plasma with frequency in the Terahertz region.
 
\end{abstract}

\date{\today}

\maketitle
Nonlinear phenomena are widely used in the design of Terahertz (0.1 THz to 10 THz, less investigated in the whole electromagnetic spectrum and termed as the ``THz gap") laser sources. For example, up-conversion from microwave GHz using the step recovery diode, is based on the nonlinear sum-frequency mixing. While down-conversion from the visible and ultraviolet is based on the nonlinear difference-frequency mixing, also used to detect electromagnetic radiations by the electro-optic sampling \cite{Heinz1,Zhangxc}. The efficiency of THz-pulse generation was further improved by the velocity matching from the pulse front tilting \cite{Heb}. 

The requirement of broadband THz sources and detectors suggests the use of materials with second-order optical nonlinearities. Recently, high-efficiency solar energy harvesting was discovered in non-centrosymmetric crystal structures such as Bismuth ferrites, perovskite oxides, ferroelectric materials and topological insulators \cite{Rappe1,Rappe2,Grinberg,Nie,Shi,Quile,Rappe3,Cook}. The nonlinear photovoltaics in non-centrosymmetric crystals is also termed as shift current photovoltaics \cite{Rappe1,Cook}. The nonlinear current is connected to a shift vector, which is the subtraction of the Berry connection of the conduction and the valence band. The Berry connection is an expectation value of the position operator, and its curl is the Berry curvature, related to topological quantities of the material.

A typical second-order nonlinear process was first demonstrated by projecting a laser beam with frequency $\omega$ through crystalline quartz \cite{Franken} to generate a laser beam with frequency $2\omega$. Later on the second-harmonic generation (SHG) was found in other materials (e.g., silicon surfaces) \cite{Heinz} with broken inversion symmetry. Nonlinear optical analogs, including SHG, have also been studied recently in various contexts, including Josephson plasma waves \cite{Savel} and cavity quantum electrodynamics \cite{Kockum,Kockum1,Stassi,Gu}. Theoretically, the nonlinear susceptibility tensor was used to describe the nonlinear electrical current in response to the external electric field, in semiconductors \cite{Sipe}, and single-layer graphene \cite{Mikhailov,Daria} with oblique incidence of radiation on the 2D electron layer. In the long-wavelength limit (the wave vector $\mathbf{q}\rightarrow0$, normal incidence), the nonlinear optical conductivity vanishes for materials with inversion symmetry.

Nonlinear susceptibility $\chi$ appears in the relation of nonlinear polarization $P$ and the electric field $E$, $P_{\alpha}(\omega_{1}+\omega_{2})=\chi_{\alpha\beta\gamma}(\omega_{1}+\omega_{2})E_{\beta}(\omega_{1})E_{\gamma}(\omega_{2})$. The electric current $J(t)$ and the polarization is connected by $J(t)=dP(t)/dt$. The nonlinear conductivity $\tilde{\sigma}$ connects the electric current and the electric field, $J_{\alpha}(\omega_{1}+\omega_{2})=\tilde{\sigma}_{\alpha\beta\gamma}(\omega_{1}+\omega_{2})E_{\beta}(\omega_{1})E_{\gamma}(\omega_{2})$. It is obtained from two approaches, the velocity operator approach \cite{Bloemb}, Eq.~(2-48) and Eq.~(2-49); and the position operator approach \cite{Rappe1,Rappe2,Rappe3}. To establish the connection between velocity-operator approach and postion-operator approach, we start from a two-band model. For example, the Hamiltonian $H_{0}$, eigenstates $u(\mathbf{k},s)$
and the corresponding eigenvalues $\varepsilon_{s}(\mathbf{k})$ are
defined in the equation $H_{0}u(\mathbf{k},s)=\varepsilon_{s}(\mathbf{k})u(\mathbf{k},s),$
where $s$ is the band index and $\mathbf{k}$ the momentum. The $x$ component of the velocity
operator matrix element is defined as $v_{s's}^{x}=\langle s'|\frac{\partial H_{0}}{\partial k_{x}}|s\rangle$
where $u(\mathbf{k},s)$ is denoted by $|s\rangle$. The $x$ component of the position
operator matrix element is defined as $r_{s's}^{x}=i\langle s'|\frac{\partial}{\partial k_{x}}|s\rangle$.
Take a second derivative of the Hamiltonian matrix elements, we obtained,
\[
\frac{\partial}{\partial k_{\alpha}}\frac{\partial}{\partial k_{\beta}}\langle s'|H_{0}|s\rangle=\delta_{ss'}\frac{\partial}{\partial k_{\alpha}}\frac{\partial}{\partial k_{\beta}}\varepsilon_{s}(\mathbf{k})\,.
\]
The $x$ component of the shift vector operator $\dot{R}$ is defined as
\[
\dot{R}_{s's}^{x}=\frac{i\partial}{\partial k_{x}}+\mathcal{L}_{s'}^x-\mathcal{L}_s^x\,,
\]
where $\mathcal{L}_{n}(\mathbf{k})=i\left\langle n\left|\nabla_{\mathbf{k}}\right| n\right\rangle\equiv(\mathcal{L}_n^x,\mathcal{L}_n^y,\mathcal{L}_n^z)$ is the Berry connection. Note that our shift vector operator $\dot{R}$ is different from the shift vector \cite{Rappe1,Rappe2,Fre,Jun} $R_{n m}^{x, y}=\frac{\partial \phi_{n m}^{y}}{\partial k_{x}}+\mathcal{L}_{n}^x-\mathcal{L}_{m}^x$, where $\phi_{n m}^{y}$ is the phase of the optical-transition matrix element $r_{n m}^{y}=\left|r_{n m}^{y}\right| e^{-i \phi_{n m}^{y}}$. 
While the shift vector $R_{n m}^{x, y}$ is a number itself, the shift vector operator $\dot{R}_{s's}^{x}$ contains a derivative operator which will take a partial derivative of the function following it. Although the shift vector operator $\dot{R}$ and the shift vector $R$ look very similar, the shift current \cite{Rappe1, Rappe2} is very different from the second-order nonlinear optical current, as we will see clearly in the discussions of Table II. 

If the particle-hole symmetry is preserved, we have $\varepsilon_{+}(\mathbf{k})=-\varepsilon_{-}(\mathbf{k})=E$
and $v_{++}^{x}=-v_{--}^{x}$. As shown in the supplementary material\cite{Supp},
velocity matrix elements are connected with position matrix elements
and shift vector operator in the following equations, for $\tilde{\sigma}_{xxx}$
the equation is 
\begin{equation}
v_{+-}^{x}v_{++}^{x}v_{-+}^{x}=2iE^{3}r_{+-}^{x}\dot{R}_{-+}^{x}r_{-+}^{x}+iE^{2}r_{+-}^{x}\tilde{m}_{-+}^{xx}\,,\label{vr}
\end{equation}
where $\tilde{m}_{ss'}^{\alpha\beta}=\langle s|\frac{\partial^{2}H_{0}}{\partial k_{\alpha}\partial k_{\beta}}|s'\rangle$ is proportional to the inverse of the effective mass defined from the band curvature. For a flat band, the second term in Eq.~[\ref{vr}] vanishes, the product of shift vector operator and position operator is equivalent to the product of velocity operator. However, this conclusion is only valid for the mixture of inter-band and intra-band velocity matrix elements such as $v_{+-}^{x}v_{++}^{x}v_{-+}^{x}$. For the pure intra-band velocity matrix elements $v_{++}^{x}v_{++}^{x}v_{++}^{x}$ such a relation does not exist.

For the linear conductivity, it is well known a trace of velocity
operators and Green's functions is connected to the product of velocity
matrix elements. We now define the nonlinear conductivity as a trace
of velocity operators and Green's functions \cite{Nori_Li}; the momentum
and imaginary frequency parameters in each Green's function is set
by using a triangle Feynman diagram, 
\begin{eqnarray}
&&\tilde{\sigma}_{\alpha\beta\gamma}(\omega_{a},\omega_{b})=\frac{i e^{3}}{\omega_{a}\omega_{b}}\int d\mathbf{k} T\sum_{l}\mathrm{Tr}\langle v_{\alpha}\widehat{G}(\mathbf{k,}i\omega_{l})v_{\beta}\times\notag \\
&&\widehat{G}(\mathbf{k,}i\omega_{l}+i\omega_{n1})v_{\gamma}\widehat{G}(\mathbf{k,}i\omega_{l}-i\omega_{n2})\rangle_{i\omega_{n1(2)}\rightarrow\omega_{a(b)}+i\delta}\,.
\end{eqnarray}

Here $e$ is the charge of the electron, $T$ is the temperature with $\omega_{n1,2}$ = 2n$\pi$$T$ the Boson Matsubara frequencies, $\omega_l = (2l+1)$$\pi$$T$ the Fermion Matsubara frequencies, $n$ and $l$ are integers. $T_r$  is a trace,   $\int d\mathbf{k} =\int dk_x/$(2$\pi$) for one-dimension(1D) and  $\int d\mathbf{k} =\int dk_x dk_y/(4{\pi^2}$) for two-dimension(2D). For simplicity we set $\hbar$ = 1, $c$ = 1 and $k_B$ = 1. To obtain the nonlinear conductivity, which is a real frequency quantity, we need to make an analytic continuation from imaginary $i\omega_n$ to real $\omega$ and $\delta$ is infinitesimal. 
In the paper \cite{Nori_Li}, the inter-band nonlinear optical conductivity of the 2D hexagonal warping model is discussed. We emphasize that the intra-band nonlinear effect is significant in materials with abrupt changes in velocity, analogous to the abrupt change of current in a step recovery diode. We use both a 1D tilted Dirac cone model and a 2D hexagonal warping model to illustrate the intra-band effect on second-order nonlinear current and high harmonic generation. 

The matrix Green’s function $\widehat{G}(\mathbf{k,}i\omega_{l})$ could be obtained from the spectral function, which is directly measured in the angular resolved photoemission spectroscopy (ARPES) experiment; or from theoretical evaluation of the self energy  $\sum(\mathbf{k,}i\omega_{l})$ , which has contributions from electron-electron, electron-phonon scattering or disorder scattering. Theoretically the Green's function is obtained from $\widehat{G}^{-1}(\mathbf{k,}i\omega_{l})=\widehat{G}_{0}^{-1}(\mathbf{k,}i\omega_{l})+\sum(\mathbf{k,}i\omega_{l})$,
where $\widehat{G}_{0}(\mathbf{k,}i\omega_{l})$ is the free electron Green's function satisfying $\widehat{G}_{0}^{-1}(\mathbf{k,}i\omega_{l})=i\omega_{l}-H_{0}(\mathbf{k})$.
In general, $\widehat{G}(\mathbf{k},i\omega_{l})$
is connected to the matrix spectral function
$\widehat{A}(\mathbf{k},\omega)$ as
\begin{equation}
\widehat{G}(\mathbf{k},i\omega_{l})=\int_{-\infty}^{\infty}\frac{d\omega}{2\pi}\frac{\widehat{A}(\mathbf{k},\omega)}{i\omega_{l}-\omega}\,,\label{ImG}
\end{equation}
then the conductivity becomes 
\begin{equation}
\tilde{\sigma}_{\alpha\beta\gamma} = \frac{i e^{3}}{\omega_{a}\omega_{b}}\int A(\mathbf{k}) B(\omega_{n1,2})_{i\omega_{n1(2)}\rightarrow\omega_{a(b)}+i\delta}
\end{equation}
here the integrals are $\int =\int d\mathbf{k} \int d\omega_{1,2,3}$ where $\int d\omega_{1,2,3}=\int_{-\infty}^{\infty}\frac{d\omega_{1}}{2\pi}\int_{-\infty}^{\infty}\frac{d\omega_{2}}{2\pi}\int_{-\infty}^{\infty}\frac{d\omega_{3}}{2\pi}$.
We have defined two functions
\begin{equation}
A(\mathbf{k})=\mathrm{Tr}\langle v_{\alpha}\widehat{A}(\mathbf{k,}\omega_{1})v_{\beta}\widehat{A}(\mathbf{k,}\omega_{2})v_{\gamma}\widehat{A}(\mathbf{k,}\omega_{3})\rangle
\end{equation} which does not depend on $\omega_{n1,2}$ and
\begin{equation}
B(\omega_{n1,2})=T\sum_{l}\frac{1}{i\omega_{l}-\omega_{1}}\frac{1}{i\omega_{l}+i\omega_{n1}-\omega_{2}}\frac{1}{i\omega_{l}-i\omega_{n2}-\omega_{3}}
\end{equation}
which does not depend on $\mathbf{k}$. Both of them depend on $\omega_{1,2,3}$. Performing the sum over Matsubara frequencies, with details in the supplementary, we obtain 
\begin{eqnarray}
&&B(\omega_{n1,2})=\frac{1}{i\omega_{n1}-\omega_{2}+\omega_{1}}\notag\\
&\times&\Big[\frac{f(\omega_{3})-f(\omega_{1})}{i\omega_{n2}+\omega_{3}-\omega_{1}}+\frac{f(\omega_{2})-f(\omega_{3})}{i\omega_{n1}+i\omega_{n2}+\omega_{3}-\omega_{2}}\Big]\,.
\end{eqnarray} 
Here $f(x)=1/[\exp(x/T-\mu/T)+1]$ is the Fermi-Dirac distribution function obtained from the sum over the internal Fermion Matsubara frequencies $\omega_{l}$. We try to avoid using $\mathbf{k}\rightarrow\mathbf{k}+e\mathbf{k}_A/\hbar$, where $\mathbf{k}_A$ is the electric vector potential, as this technique requires the treatment of the diamagnetic and paramagnetic terms \cite{VPGusynin}. Without separating into diamagnetic and paramagnetic terms, we recover the same sum rule \cite{LiZhou}, which establishes the validity of our method. In the discussions of high harmonic generation, we have to use the technique $\mathbf{k}\rightarrow\mathbf{k}+e\mathbf{k}_A/\hbar$ to obtain a quick estimate of the intra-band contribution to the high harmonic generation.

From Eq.~(\ref{ImG}), we know $\widehat{A}(\mathbf{k,}\omega)=-2\rm{Im} \widehat{G}(\mathbf{k},\omega)$. For a two-band model, the free electron spectral function consists of delta-functions, 
\[
\widehat{A}(\mathbf{k,}\omega)=2\pi\delta(\omega+\mu-\varepsilon_{+})|+\rangle\langle+|+2\pi\delta(\omega+\mu-\varepsilon_{-})|-\rangle\langle-|
\]
For  $\alpha=x$ or $y$, $\beta=x$ and $\gamma=x$ the trace in $A(\mathbf{k})$ is carried out 
\begin{eqnarray}
&&A(\mathbf{k})=8\pi^3\Big[v_{++}^{x(y)}v_{++}^{x}v_{++}^{x}\delta_{+++}+v_{+-}^{x(y)}v_{--}^{x}v_{-+}^{x}\delta_{--+} \notag\\
&&+v_{++}^{x(y)}v_{+-}^{x}v_{-+}^{x}\delta_{+-+}+v_{+-}^{x(y)}v_{-+}^{x}v_{++}^{x}\delta_{-++}\notag\\
&&+v_{-+}^{x(y)}v_{++}^{x}v_{+-}^{x}\delta_{++-}+v_{--}^{x(y)}v_{--}^{x}v_{--}^{x}\delta_{---}\notag\\
&&+v_{-+}^{x(y)}v_{+-}^{x}v_{--}^{x}\delta_{+--}+v_{--}^{x(y)}v_{-+}^{x}v_{+-}^{x}\delta_{-+-}\Big]
\end{eqnarray}
with the definition $\delta_{s_1s_2s_3}=$
\begin{equation*}
\delta(\omega_{1}+\mu-\varepsilon_{s_1})\delta(\omega_{2}+\mu-\varepsilon_{s_2})\delta(\omega_{3}+\mu-\varepsilon_{s_3})
\end{equation*}
Now we separate the inter-band and intra-band process, $\tilde{\sigma}_{xxx}=\tilde{\sigma}^{\textrm{inter}}_{xxx}+\tilde{\sigma}^{\textrm{intra}}_{xxx}$, and $A(\mathbf{k})=A^{\textrm{inter}}(\mathbf{k})+A^{\textrm{intra}}(\mathbf{k})$. Collecting all the inter-band contributions for $\tilde{\sigma}^{\textrm{inter}}_{xxx}$ we have
\begin{eqnarray}
A^{\textrm{inter}}(\mathbf{k})&=&8\pi^3v_{+-}^{x}v_{-+}^{x}\Big[v_{++}^{x}(\delta_{+-+}+\delta_{-++}+\delta_{++-})\notag\\
&+&v_{--}^{x}(\delta_{+--}+\delta_{-+-}+\delta_{--+})\Big]
\end{eqnarray}
Collecting all the intra-band contributions for $\tilde{\sigma}^{\textrm{intra}}_{xxx}$ we have
\begin{equation}
A^{\textrm{intra}}(\mathbf{k})=8\pi^3\Big[(v_{++}^{x})^3\delta_{+++}+(v_{--}^{x})^3\delta_{---}\Big]
\end{equation}

In the paper \cite{Sipe1}, the second-order nonlinear optical conductivity is separated into four terms: a pure intraband term, a pure interband term, and two mixing terms. We have all these terms from the velocity-operator approach. However, after the spectral expansion of the Green’s function (Eq.~\ref{ImG}), we separate the momentum integration $A(\mathbf{k})$ and the frequency integration $B(\omega)$. In $A(\mathbf{k})$, we have the pure intraband product of velocity matrix elements and interband-intraband mixing of velocity matrix elements. This does not mean the pure interband term is missing; it is contained in the velocity matrix element mixing terms. In the structure of $B(\omega)$ one can see clearly the pure interband term. 

\begin{figure}
    \centering
    \includegraphics[width=0.43\textwidth,height=0.37\textwidth]{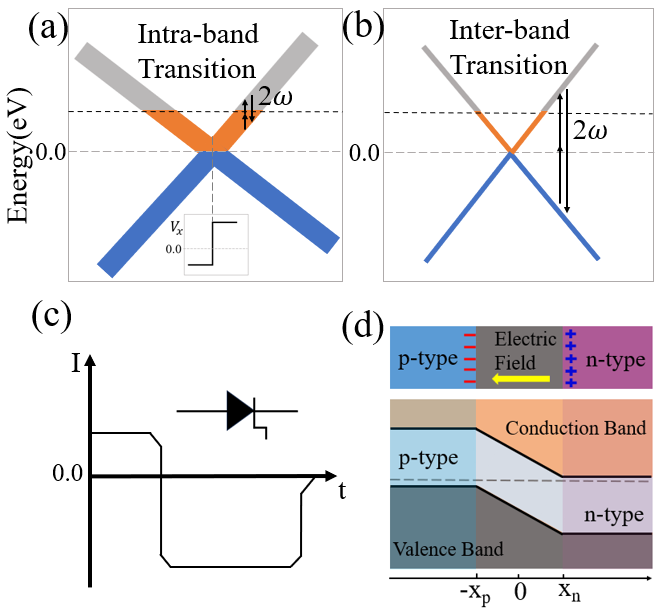 }
    \caption{Schematic of (a) intra-band and (b) inter-band  transition \cite{Jiang}. In the inset of (a), the velocity changes abruptly as a function of k. (c) Current and time relationship of the step recovery diode. The current changes abruptly as a function of t. (d) Schematic of a pn-junction and its energy band. From Poisson's equation, the redistribution of the electric charge concentration in the pn-junction is connected to the strong internal electric field in materials, which is important to realize asymmetric energy as depicted in (a).   }
    \label{fig:FIG. 1}
\end{figure}

For the interband nonlinear conductivity, the delta functions $\delta_{s_1s_2s_3}$ could be used to carry out the integral $\int d\omega_{1,2,3}$. For nonequal frequencies $\omega_{a}$ and $\omega_{b}$, we have
\begin{eqnarray}
&&\tilde{\sigma}_{xxx}^{\textrm{inter}}=\frac{ie^{3}}{\omega_{a}\omega_{b}}\int d\mathbf{k} \sum_{s=\pm}\Big[\sum_{n=a,b}\frac{f(\varepsilon_{-})-f(\varepsilon_{+})}{\omega_{a}+\omega_{b}+2i\delta+sp\varepsilon_{\Delta}} \notag\\
&\times&\frac{sv_{+-}^{x}v_{-+}^{x}v^{x}_{ss}}{\omega_{n}+i\delta+sp\varepsilon_{\Delta}}+\frac{f(\varepsilon_{+})-f(\varepsilon_{-})}{\omega_{b}+i\delta+s\varepsilon_{\Delta}}\frac{sv_{+-}^{x}v_{-+}^{x}v^{x}_{ss}}{\omega_{a}+i\delta-s\varepsilon_{\Delta}}\Big]
\end{eqnarray}
where $\varepsilon_{\Delta}=\varepsilon_{+}-\varepsilon_{-}$, $p=+1$ for $n=a$ and $p=-1$ for $n=b$. Note that if the inversion symmetry was not broken, i.e. $v^{x}(\mathbf{k})v^{x}(\mathbf{k})v^{x}(\mathbf{k})=-v^{x}(-\mathbf{k})v^{x}(-\mathbf{k})v^{x}(-\mathbf{k})$, the integrand is an odd function of $\mathbf{k}$ so the integration in momentum space vanishes. For the intra-band nonlinear conductivity, the integral $\int d\omega_{1,2,3}$ is not integrated out and we use the same notation  $\int =\int d\mathbf{k} \int d\omega_{1,2,3}$,
\begin{eqnarray}
&& \tilde{\sigma}_{xxx}^{\textrm{intra}}=\frac{ie^{3}}{\omega_{a}\omega_{b}}\int \frac{A^{\textrm{intra}}(\mathbf{k})}{\omega_a+i\delta-\omega_{2}+\omega_{1}}\times\notag\\
&& \Big[\frac{f(\omega_{3})-f(\omega_{1})}{\omega_b+i\delta+\omega_{3}-\omega_{1}}+\frac{f(\omega_{2})-f(\omega_{3})}{\omega_a+\omega_b+2i\delta+\omega_{3}-\omega_{2}}\Big]
\end{eqnarray}

Now we are ready to discuss two limit cases $\omega_{a}=\omega_{b}=\omega$ and $\omega_{a}=-\omega_{b}=\omega$, for the inter-band nonlinear optical conductivity we have
\begin{eqnarray}
&&\tilde{\sigma}_{xxx}^{\textrm{inter}}(\omega,\omega)=\frac{ie^{3}}{\omega^{2}}\int d\mathbf{k}\frac{v_{+-}^{x}v_{-+}^{x}v^{x}_{\Delta}}{\varepsilon_{\Delta}}\times\notag\\
&&\sum_{s=\pm}\Big[\frac{0.5s[f(\varepsilon_{+})-f(\varepsilon_{-})]}{\omega+i\delta+s\varepsilon_{\Delta}}+\frac{s[f(\varepsilon_{+})-f(\varepsilon_{-})]}{\omega+i\delta-0.5s\varepsilon_{\Delta}}\Big].
\end{eqnarray}
where $v^{x}_{\Delta}=v_{++}^{x}-v_{--}^{x}$. However for the $\omega_{a}=-\omega_{b}=\omega$ case, the nonlinear conductivity is a pure imaginary number, as $\tilde{\sigma}_{xxx}^{\textrm{inter}}(\omega,-\omega)=$
\begin{eqnarray}
&&\frac{ie^{3}}{\omega^{2}}\int d\mathbf{k} \sum_{s=\pm} s v_{+-}^{x}v_{-+}^{x}v^{x}_{ss} \Bigg[\frac{f(\varepsilon_{+})-f(\varepsilon_{-})}{(\omega-s\varepsilon_{\Delta})^{2}+\delta^{2}} \notag\\
&&+\frac{2(\varepsilon^2_{\Delta}+s\omega\varepsilon_{\Delta}-2\delta^2)[f(\varepsilon_{+})-f(\varepsilon_{-})]}{(\varepsilon^2_{\Delta}+s\omega\varepsilon_{\Delta}-2\delta^2)^2+\delta^2(3s\varepsilon_{\Delta}+2\omega)^2}\Bigg]\,.
\end{eqnarray}
This breaks the Kramers-Kronig relation for the inter-band nonlinear optical conductivity. For the intra-band part, we do not use the free-electron spectral function, but keep the self-energy from the impurity scattering. So the delta function becomes a broadened Lorentzian $\delta(\omega_{1}+\mu -\varepsilon_{\pm})\rightarrow(-1/\pi)\rm{Im}[1/(\omega_{1}+i\delta+\mu -\varepsilon_{\pm})]=S_{\pm}(\omega_{1})$, for the  $\omega_{a}=\omega_{b}=\omega$ case we have
\begin{eqnarray}
&&\tilde{\sigma}_{xxx}^{\textrm{intra}}(\omega,\omega) =  \frac{ie^{3}}{\omega^{2}} \int  \frac{S(\omega_1,\omega_2,\omega_3)}{\omega_{3}+\omega_{2}-2\omega_{1}}\Big[\frac{f(\omega_{2})-f(\omega_{1})}{\omega+i\delta-\omega_{2}+\omega_{1}}  \notag\\
&&+\frac{f(\omega_{1})-f(\omega_{3})}{\omega+i\delta+\omega_{3}-\omega_{1}}+\frac{f(\omega_{3})-f(\omega_{2})}{\omega+i\delta+0.5\omega_{3}-0.5\omega_{2}}\Big]\,,
 \end{eqnarray}
where $S(\omega_1,\omega_2,\omega_3)=8\pi^3[(v_{++}^{x})^3 S_{+}(\omega_1)S_{+}(\omega_2)S_{+}(\omega_3)+(v_{--}^{x})^3 S_{-}(\omega_1)S_{-}(\omega_2)S_{-}(\omega_3)]$. For the $\omega_{a}=-\omega_{b}=\omega$ case, $\tilde{\sigma}^{\textrm{intra}}_{xxx}(\omega,-\omega) = $
\begin{eqnarray}
&&\frac{-ie^{3}}{\omega^{2}}\int 
\frac{S(\omega_1,\omega_2,\omega_3)}{\omega_{3}+\omega_{2}-2\omega-2\omega_{1}}\Big[\frac{f(\omega_{3})-f(\omega_{2})}{i\delta+0.5(\omega_{3}-\omega_{2})}\notag\\
&& +\frac{f(\omega_{1})-f(\omega_{3})}{-\omega+i\delta+\omega_{3}-\omega_{1}}+\frac{f(\omega_{2})-f(\omega_{1})}{\omega+i\delta-\omega_{2}+\omega_{1}}\Big]\label{w-w}
\end{eqnarray}
If we focus on the real part of the nonlinear conductivity, the equation is further simplified, we reduce the three-dimenion integration over the frequency to a two-dimension one, which we numerically sovled in \cite{Li22}
\begin{eqnarray}
&\ \rm{Re}\tilde{\sigma}^{\textrm{intra}}_{xxx}(\omega,\omega) = \frac{e^{3}}{2\omega^{2}}\int d\mathbf{k} \int d\omega_{1,2} \Big[ S(\omega_1,\omega_2,\omega_1+\omega)\times\notag\\
&\ \frac{f(\omega_{1}+\omega)-f(\omega_1)}{\omega_{2}-\omega_{1}+\omega}+S(\omega_1,\omega_2,\omega_1-\omega)\frac{f(\omega_{1})-f(\omega_1-\omega)}{\omega_{2}-\omega_{1}-\omega}\notag\\
&\ +S(\omega_1,\omega_2,\omega_2-2\omega)\frac{f(\omega_{2}-2\omega)-f(\omega_2)}{\omega_{2}-\omega-\omega_1}\Big]\label{Real}
\end{eqnarray}
From Eq.~(\ref{w-w}) we find $\rm{Re}\tilde{\sigma}^{\textrm{intra}}_{xxx}(\omega,-\omega)=0$, so both the inter-band and intra-band of $\tilde{\sigma}_{xxx}(\omega,-\omega)$ is a pure imaginary number, this breaks the Kramers-Kronig relation.

In Fig.~\ref{fig:FIG. 1} (a),  we propose a scenario in which the intra-band process is nonzero, with a band structure  asymmetric in the Brillouin zone. In Fig.~\ref{fig:FIG. 1} (b), due to the imaginary part of the inter-band velocity matrix element being an even function of k, the inter-band nonlinear optical conductivity is nonzero \cite{Jiang}, although the band is symmetric in the momentum space. In Fig.~\ref{fig:FIG. 1} (c), we describe the I-t curve of the step recovery diode, the abrupt change of the electric current is an analogy of the abrupt change of the velocity. In Fig.~\ref{fig:FIG. 1} (d), we present the energy curve of a pn-junction, which is asymmetric in the real space and also in the momentum space (with a Fourier transform). 

Firstly, we look at a 1D tilted Dirac cone model which contains the abrupt change in velocity and also breaks the inversion symmetry,
\begin{equation}
H_{0}=-2\gamma k_{x}+\hbar v_{\mathrm{F}}k_{x}\sigma_{z}\label{1DD}
\end{equation}
The energy of the two bands are $\varepsilon_{\pm}(k_x)=-2\gamma k_{x}\pm\hbar v_{\mathrm{F}}|k_{x}|$. The velocity operator $v_x=-2\gamma/\hbar+ v_{\mathrm{F}}\sigma_{z}$, the velocity matrix elements are $v_{++}^{x}=-2\gamma/\hbar+ v_{\mathrm{F}}\rm{sgn}(k_x)$, $v_{--}^{x}=-2\gamma/\hbar-v_{\mathrm{F}}\rm{sgn}(k_x)$ and $v_{+-}^{x}=v_{-+}^{x}=v_{\mathrm{F}}\delta(k_x)$. Typical parameters $\hbar v_{\mathrm{F}}=$ 1 eV$\cdot\mathrm{\mathring{A}}$ and $\gamma$ = 1 eV$\cdot\mathrm{\mathring{A}}$. In three dimensional (3D) type-II Weyl semimetal (e.g. WTe$_{2}$), pairs of tilted Dirac cones were found. The search for a single tilted Dirac cone is of great interest. 

Then we consider a 2D hexagonal warping model \cite{Fu,Li2,Li1}, 
\begin{equation}
H_{0}=\hbar v_{\mathrm{F}}(k_{x}\sigma_{y}-k_{y}\sigma_{x})+\frac{\lambda}{2}(k_{+}^{3}+k_{-}^{3})\sigma_{z}+M\sigma_{z}\,,
\end{equation}
this model has been used to describe the surface states of a 3D topological insulator(TI) and also in ferroelectric materials. The Fermi velocity $\hbar v_{\mathrm{F}}$= 2.55 $\rm{eV}\cdot\mathrm{\mathring{A}}$, the hexagonal warping $\lambda$=250 $\rm{eV}\cdot\mathrm{\mathring{A}}^{3}$, $\sigma_{x}$, $\sigma_{y}$, $\sigma_{z}$ are Pauli matrices and $k_{\pm}=k_{x}\pm ik_{y}$. M is the gap parameter for a TI in proximity to magnetic impurities, e.g. in $\rm{Bi}_{2}\rm{Se}_{3}$
\cite{Xu1,Xu2,Chen1,Hasan,Qi1,Moore,Hsieh1,Chen,Hsieh2} and $\rm{Cr}_{x}(\rm{Bi}_{1-y}\rm{Sb}_{y})_{2-x}\rm{Te}_{3}$ \cite{Tokura,Yasuda}. The eigenvalue of this model is $\varepsilon_{\pm}=\pm\sqrt{\hbar^2v_{\mathrm{F}}^{2}k^{2}+ \varepsilon_{\lambda M}^{2}}$. The velocity $v_{++}^{x}=[\hbar k_{x}v_{\mathrm{F}}^2+3\lambda(k_{x}^2-k_{y}^2)\cdot\varepsilon_{\lambda M}/\hbar]/\varepsilon_{+}$
, here $\varepsilon_{\lambda M}=\lambda (k_{x}^3-3k_{x}k_{y}^2)+M$.

In table I, we present various symmetries (rotation, inversion and time reversal) for four typical cases of the hexagonal warping model. Note that only when both $M\neq0$ and $\lambda\neq0$, the inversion symmetry is broken, second order nonlinear conductivity is nonzero in the long-wavelength limit $\mathbf{q}\rightarrow0$. 

\begin{table}
\caption{\label{Symm}Symmetries for different cases of the 2D hexagonal
warping model.}

\begin{ruledtabular}
\begin{tabular}{l|l|l|l|l}
 & $\lambda=0$  & $\lambda\neq0$  & $\lambda\neq0$  & $\lambda=0$\tabularnewline
 & $M=0$  & $M\neq0$  & $M=0$  & $M\neq0$\tabularnewline
\hline 
Rotation  & Yes  & No  & No  & Yes\tabularnewline
\hline 
Time reversal(T)  & Yes  & No  & Yes  & No\tabularnewline
\hline 
Inversion(P)  & Yes  & No  & Yes  & Yes\tabularnewline
\end{tabular}\end{ruledtabular}

\end{table}

\begin{figure}
    \centering
    \includegraphics[width=0.5\textwidth,height=0.35\textwidth]{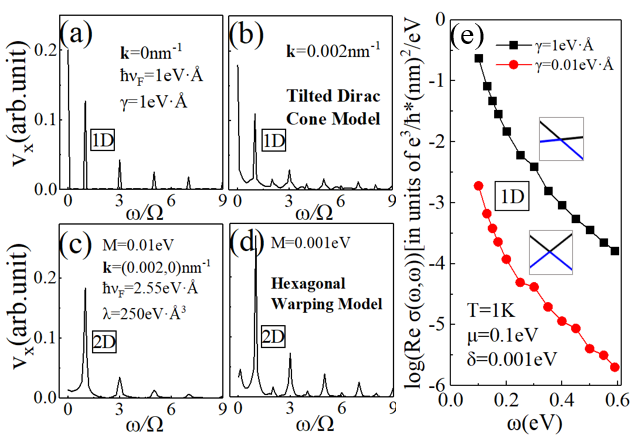 }
    \caption{ Intra-band high harmonic generation of (a, b) the 1D tilted Dirac cone model and (c, d) the 2D hexagonal warping model. The harmonics are generated from a Fourier transform of the velocity in the x direction, which changes abruptly at the $k_x=0$ point. (e) The real part of intra-band nonlinear optical conductivity obtained from Eq.~(\ref{Real}). For the 1D tilted Dirac cone model  Eq.~(\ref{1DD}), as the tiltness $\gamma$ increases, the nonlinear optical conductivity increases significantly. }
    \label{fig:FIG. 2}
\end{figure}

In Fig.~\ref{fig:FIG. 2} (a, b, c, d) we analyze the intra-band harmonic generation of 1D tilted Dirac cone model and 2D hexagonal warping model. In 1D we find the tilting gives a high peak near $\omega/\Omega=0$. In 2D, a smaller gap in the hexagonal warping model enhances the high harmonic generation. Consider an electric field pulse is applied to the material and the vector potential is $\mathbf{k}_A(t)=(E_0/\Omega) \rm{exp}(-2\rm{ln}2(t/t_0)^2)\rm{sin}(\Omega t)\hat{e}_{x}$. Here $E_0$ and $\Omega$ are the strength and the frequency of the electric field respectively, and $t_0$ is the width of the pulse. In a quasiclassical kinetic approach, we investigate the intra-band contribution to the high harmonic generation \cite{Lizy}, $v_{++}^{x}(\mathbf{k},t)=v_{++}^{x}(\mathbf{k}+e\mathbf{k}_A(t)/\hbar)$, and the fourier transform $v_{++}^{x}(\mathbf{k},\omega)=\int dt v_{++}^{x}(\mathbf{k},t) e^{i\omega t}$. We use $\mathrm{t_{0}}$ = 5 $\mathrm{ps}$, $\Omega$ = 1 $\mathrm{THz}$ and $E_{0}$ = 1.3 $\times 10^5~\rm{V}/m$.  The intra-band contribution to the real part of nonlinear optical conductivity is calculated from Eq.~(\ref{Real}). In Fig.~\ref{fig:FIG. 2} (e) we present the result of a 1D tilted Dirac cone model. The intra-band optical conductivity is two orders of magnitude larger than the normal value, and increases as the input frequency decreases. Its unit can also be expressed as $\rm{nm^2\cdot\mu A/V^2}$, when $\omega$ is 0.1 eV, the normal value of the intra-band optical conductivity is 0.001897 $\rm{e^3/\hbar\cdot(nm)^2/eV}$ = 0.461920 $\rm{nm^2\cdot\mu A/V^2}$.

In below we use density functional theory (DFT) and non-equilibrium Green's function software to calculate the electronic properties of materials from chemical elements. The electrostatic potential energy calculation of Si pn-junction, the electronic properties calculation of $\rm{B_8}$, $\rm{AlN}$ and the orbital projection band calculation for $\rm{TaAs}$ and $\rm{LiNbO_3}$ are performed by using a first-principles method based on DFT \cite{Kresse}, as implemented in the Nanodcal and DS-PAW which are programs under the Device Studio platform, for superconductor $\rm{NbN}$ we use Quantum ESPRESSO. The DS-PAW is based on the plane wave basis and the projector augmented wave (PAW) representation,the Nanodcal is a first principles computing software based on linear combined atomic orbital basis and non-equilibrium Green's function-density functional theory\cite{Blochl}. 
The Perdew-Burke-Ernzerhof (PBE) exchange-correlation energy functional within the generalized gradient approximation (GGA) are employed \cite{Perdew,Grimme}. For $\rm{NbN}$, SG15-type \cite{sg1} norm-conserving pseudo-potential is employed \cite{sg2}. The electronic iteration convergence criterion is set to $10^{-4}$ $\rm{eV}$, for $\rm{NbN}$ it is $10^{-5}$  a.u (27.21146×$10^{-5}$ $\rm{eV}$). The wave functions were expanded in plane waves up to a kinetic energy cutoff of 500 $\rm{eV}$ for $\rm{B_8}$ and $\rm{AlN}$, 300 $\rm{eV}$ for $\rm{TaAs}$ and $\rm{LiNbO_3}$, 120 Ry (1632.6 eV) for $\rm{NbN}$ and 480 Ry (6530.4 eV) for the charge density cutoff of $\rm{NbN}$, 100 Hartree (2721.1 eV) for Si pn junction. The Brillourin zone integration is obtained by using a k-point sampling mesh of $10\times13\times1$ for $\rm{B_8}$, $14\times14\times14$ for $\rm{AlN}$, $12\times12\times4$ for $\rm{TaAs}$, $10\times10\times4$ for $\rm{LiNbO_3}$, $10\times10\times2$ for hexagonal NbN, $10\times10\times10$ for cubic NbN, $11\times11\times1$ for single layer Si pn junction, $6\times11\times1$ for bilayer Si pn junction, generated according to the Gamma-centered method. We have included the contribution of spin-orbit coupling (SOC) in our calculations of $\rm{TaAs}$ and $\rm{LiNbO_3}$. The thickness of the vacuum layer was set for 20 $\mathrm{\mathring{A}}$ in calculations of $\rm{B_8}$, which can reduce the interactions between the adjacent layers. In the Si pn junction modeling, we use the virtual crystal approximation (VCA) method to realize p-type and n-type vacancy doping.
\begin{figure}
    \centering
    \includegraphics[width=3.6in]{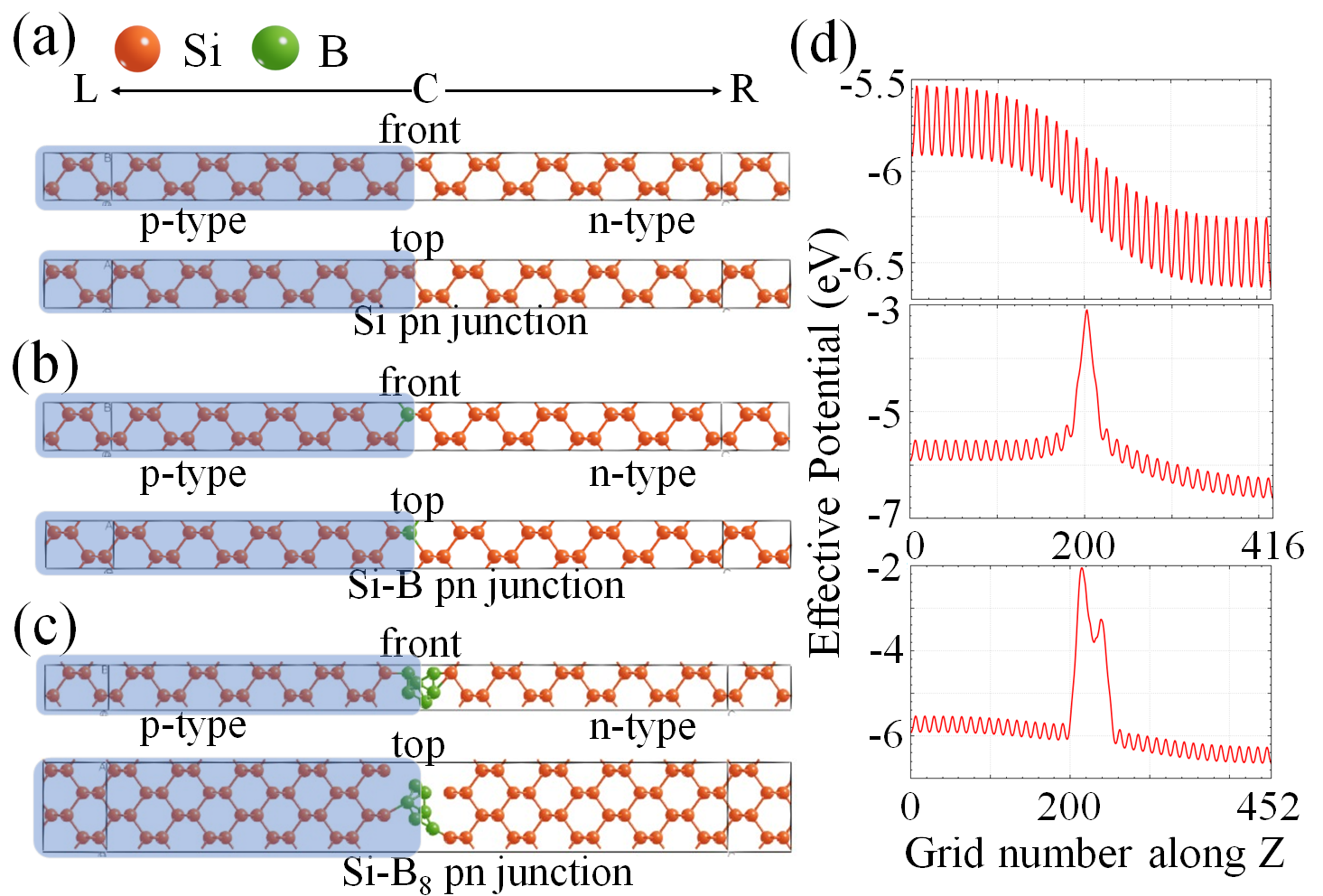}
    \caption{(a) The atomic structure of a Silicon pn-junction. The blue shaded region is the p-type Silicon. (b) The silicon pn-junction with a Boron atom replacing the Silicon atom in the middle of the junction. (c) A layer of $\rm{B_8}$ is inserted into the Silicon pn-junction. (d) The potential energy of the three devices (a), (b) and (c), from top to bottom respectively. The real space is divided into grids for simultaneous solving of the Poisson's equation and the effective equation from DFT, using the Nanodcal software. }
    \label{fig:potential}
\end{figure}

In Fig.~\ref{fig:potential} (a), we provide the top and front view of the Silicon pn junction, generated from the Device Studio platform. The device consists of 11 Silicon cells, the length of central region is 48.8754 $\mathrm{\mathring{A}}$ (9 Silicon cells), the length of left or right electrode is 5.4306 $\mathrm{\mathring{A}}$ (1 Silicon cell). In Fig.~\ref{fig:potential} (b), the Silicon pn junction is replaced by one Boron atom in the middle. In Fig.~\ref{fig:potential} (c), the Silicon pn-junction is expanded in the top view, to make it large enough for the insertion of $\rm{B_8}$ in the middle of the device.  The length of the center region increases slightly to 53.20175 $\mathrm{\mathring{A}}$. We use the VCA of Nanodcal to achieve p-type and n-type doping. The doping concentration of p-type is 0.999, and for n-type it is 1.001. The left and right poles we use in the calculation are silicon crystals (with a valence electron number of 4), and the p/n type of doping is 0.999/0.001 results in the device having a concentration of four thousandths of an electron doped. The p/n-type doping ratio is self-regulated using the VCA \cite{Bellaiche}. In Fig.~\ref{fig:potential} (d), the central region length of Si pn junction and Si-B pn junction is the same 48.8754 $\mathrm{\mathring{A}}$, and the calculated real space grid number is 416. In Si-$\rm{B_8}$ pn junction, due to the insertion of $\rm{B_8}$ cell, the central region length changes to 53.20175 $\mathrm{\mathring{A}}$, and the real space grid number is 452. Without applied external voltage on the 3 devices, the potential energy displays obvious asymmetry in the real space. Moreover, for Si-B pn-junction, there is an obvious peak at the location of B atom, and for Si-$\rm{B_8}$ pn-junction, there are two peaks at the position where $\rm{B_8}$ cell is inserted, a sharp main peak and a smaller side peak, asymmetric in the real space.

\begin{figure}
    \centering
    \includegraphics[width=3.3in]{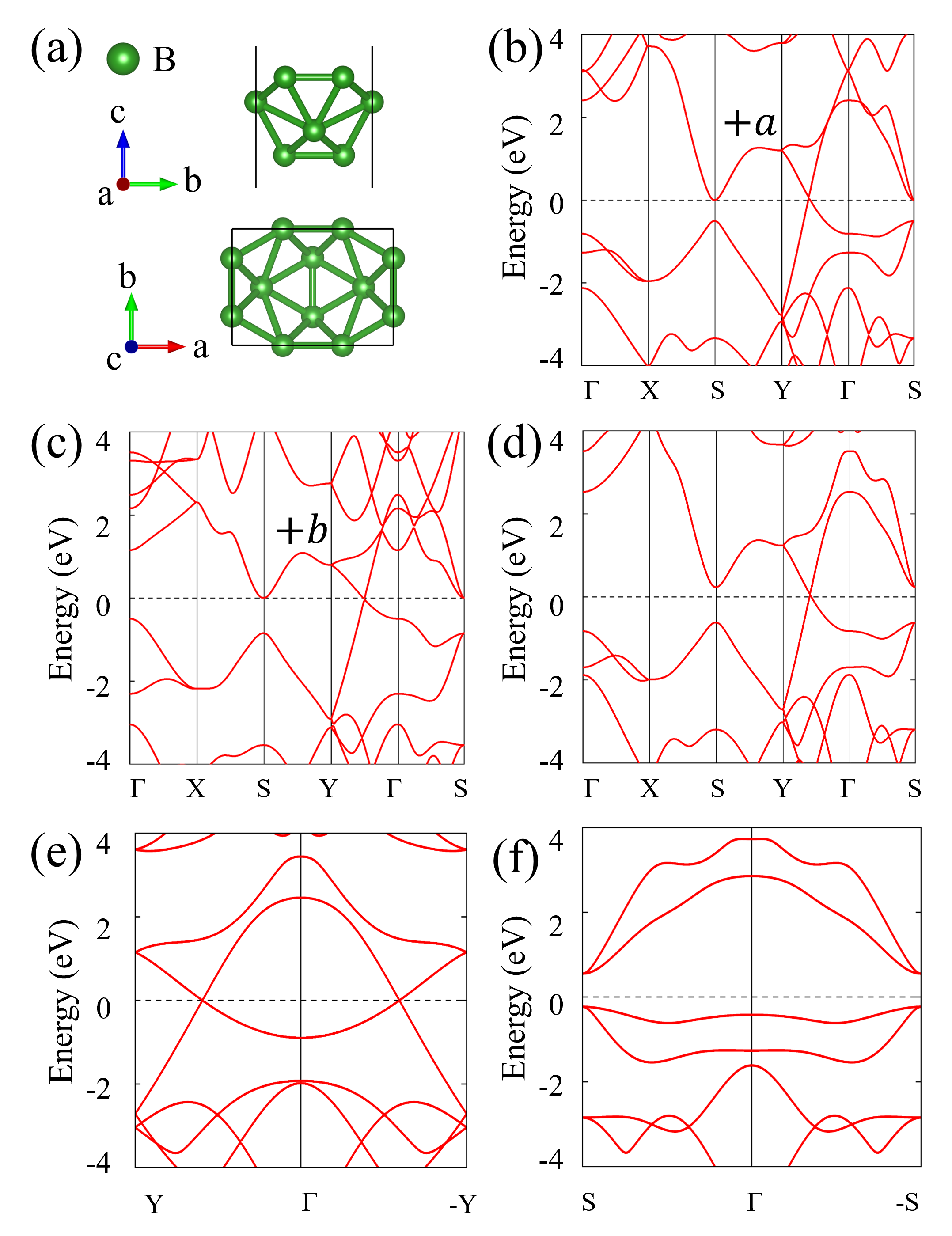}
    \caption{(a) The lattice structure of $\rm{B_8}$. (b) After stretching along the a axis, the tiltness of the Dirac cone changes slightly. (c) After stretching along the b axis. the tiltness of the Dirac cone changes significantly. The strength of the stretching is to change the original lattice constant by 10\%. (d) No stretching. (e-f) The band structure of $\rm{B_8}$ along $k$ and $-k$ direction. }
    \label{fig:FIG4}
\end{figure}

In Fig.~\ref{fig:FIG4}, we present the lattice and band structure of $\rm{B_8}$. The band gap near the high symmetry point $S$ changes significantly from 0.859 $\rm{eV}$ to 0.507 $\rm{eV}$. Stretching along the b-axis of $\rm{B_8}$, the bands near the high symmetry point $\Gamma$ shift upwards, resulting a tilt of the Dirac cone in the $\rm{Y}$-$\Gamma$ path. Stretching along the a-axis, the bands are not changed significantly. As seen in Fig.~\ref{fig:FIG. 2}, changing the tiltness of the Dirac cone significantly affects the intra-band nonlinear optical conductivity. However as seen in Fig.~\ref{fig:FIG4} (e-f), the band structure is symmetric in the whole Brillouin zone, so the tilting effects will be canceled and no intr-band nonlinear conductivity survive. Moreover, there is no SOC effect for $\rm{B_8}$, so we do not observe the Wannier-fitted Hamiltonian becoming asymmetric as it did for $\rm{TaAs}$. Therefore the inter-band nonlinear optical conductivity is also zero.

\begin{figure}
    \centering
    \includegraphics[width=3.3in]{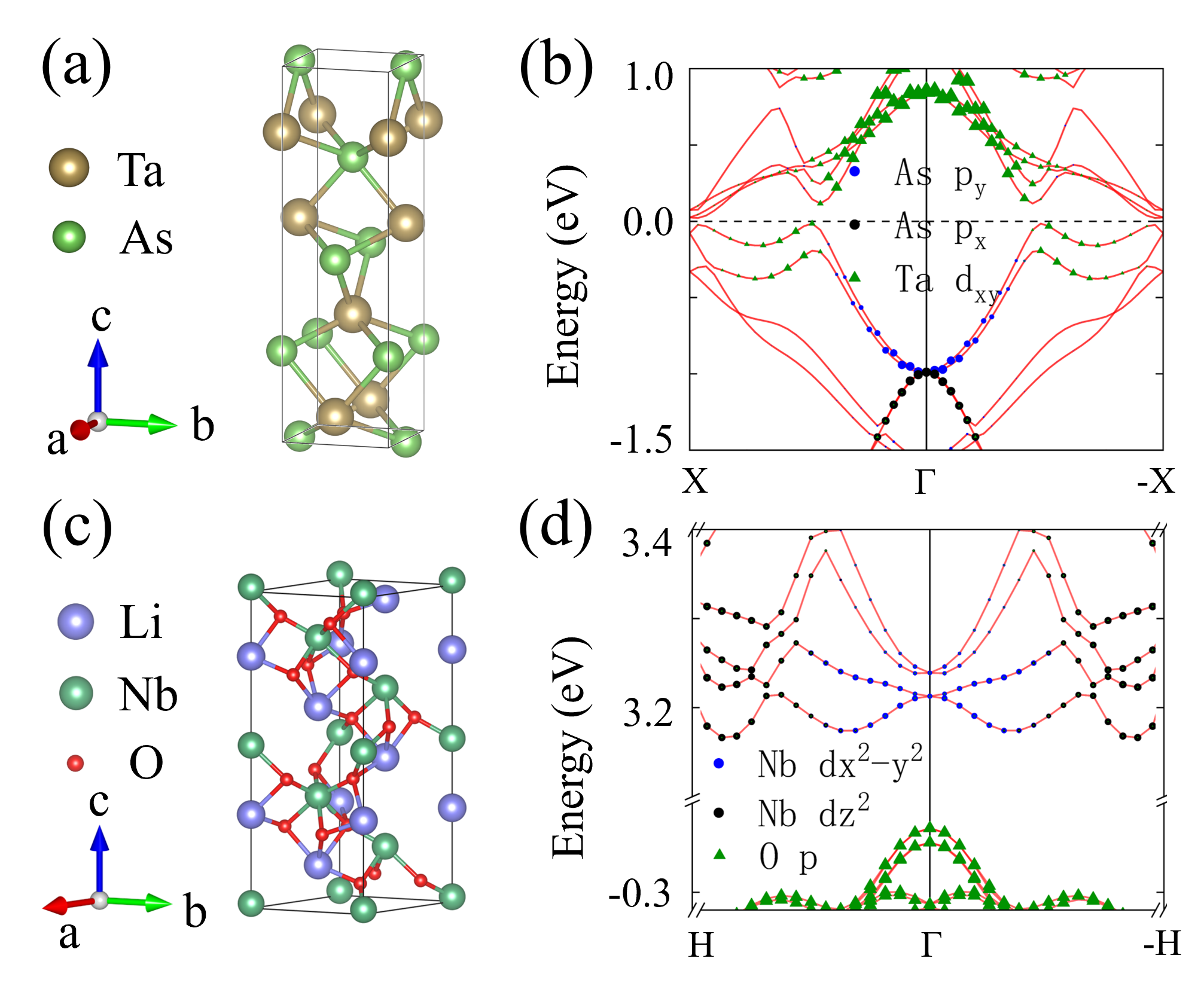}
    \caption{
    (a) The structure of $\rm{TaAs}$. 
    (b) The projected band for $\rm{TaAs}$. The high-symmetric points are X(0.0, ${\frac{1}{2}}$, 0.0), $\rm{\Gamma}$(0.0, 0.0, 0.0), -X(0.0, -${\frac{1}{2}}$, 0.0). 
    (c) The structure of $\rm{LiNbO_3}$. 
    (d) The projected band for $\rm{LiNbO_3}$. The high-symmetric points are H(${\frac{1}{3}}$, ${\frac{1}{3}}$, ${\frac{1}{2}}$), -H(${-\frac{1}{3}}$, ${-\frac{1}{3}}$, ${-\frac{1}{2}}$). }  
    \label{fig:FIG5}
\end{figure}

\begin{figure}
    \centering
    \includegraphics[width=3.5in]{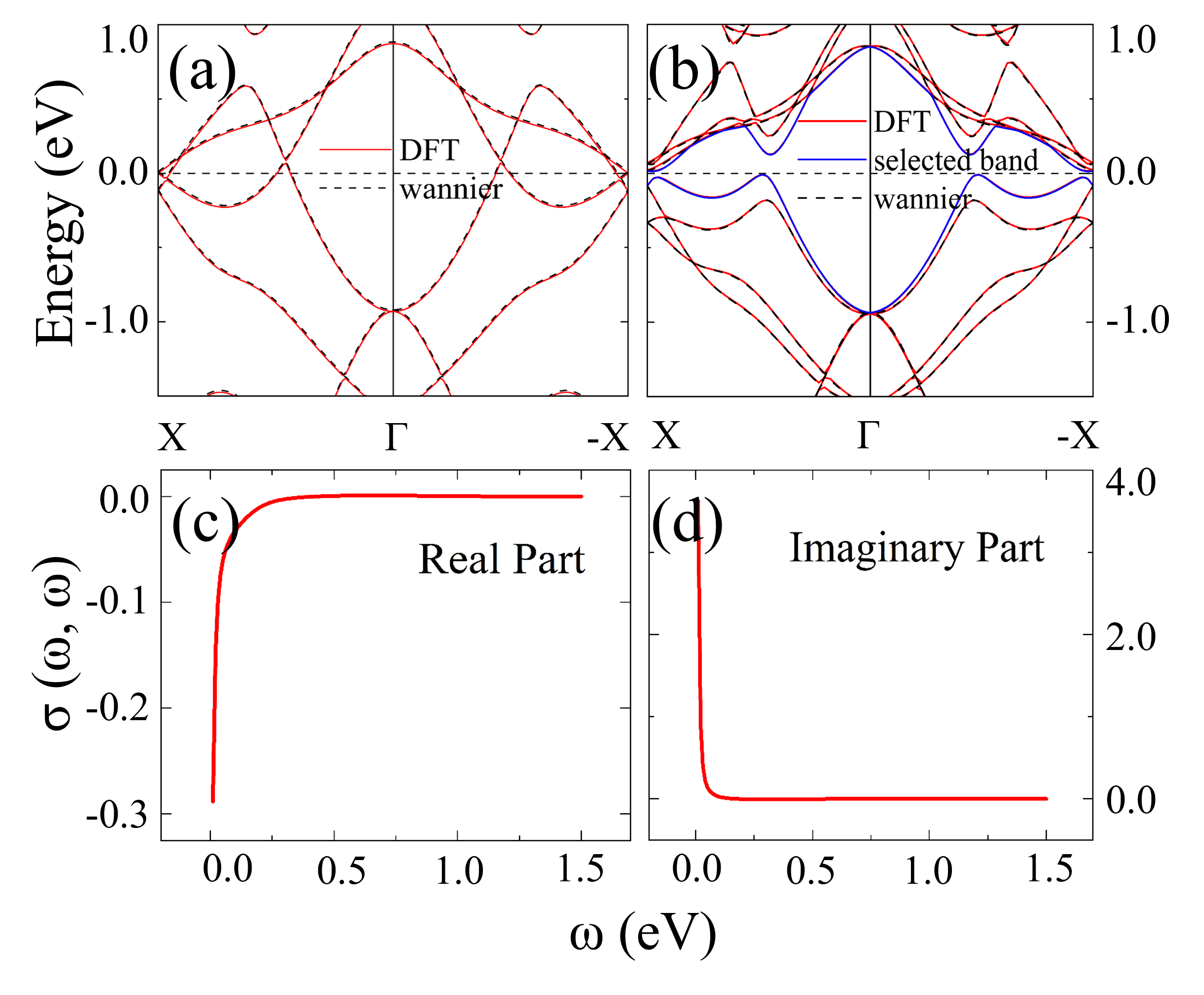}
    \caption{The Wannier fitting of the DFT band structure about (a) TaAs without SOC, and (b) TaAs with SOC. The real part (c) and the imaginary part (d) of the inter-band nonlinear optical conductivity [in units of $\rm{e^3/\hbar\cdot(nm)^2/eV}]$ of TaAs (With SOC).
    }  
    \label{fig:TaAs}
\end{figure}

In Fig.~\ref{fig:FIG5}, we present the structures and the projected bands of $\rm{TaAs}$ and $\rm{LiNbO_3}$. Around the $\Gamma$ point, the contribution to the conduction band of $\rm{TaAs}$ mainly comes from the $d_{xy}$ orbital of $\rm{Ta}$ atom,  while for $\rm{LiNbO_3}$, it mainly comes from the $d_{x^2-y^2}$ orbital of $\rm{Nb}$ atom, and switch to the $d_{z^2}$ orbital as the momentum moves away from the $\Gamma$ point. The contribution to the valence band of $\rm{TaAs}$  comes from both $p_x$ and $p_y$ orbitals of $\rm{As}$ atom, while for $\rm{LiNbO_3}$, it comes from the $p_x$, $p_y$, and $p_z$ orbitals of $\rm{O}$ atom. Moreover, $\rm{LiNbO_3}$ (also in other oxides \cite{Lei}) has a wide band gap about 3.367 $\rm{eV}$ while $\rm{TaAs}$ is a gapless semimetal.  Fig.~\ref{fig:FIG5} (d) shows only 80\% of the length on the ${\Gamma}$-H path. Although both materials are noncentrosymmetric in the real space, we do not find any asymmetry of the band structure in the Brillouin zone. So the intra-band part of the second order nonlinear optical conductivity is zero.

In Fig.~\ref{fig:TaAs} (a) and (b), we show the band structure of TaAs. DFT calculations are carried out using the Vienna Ab-initio simulation package (VASP) \cite{vasp1,vasp2}. The Wannier fitting of the DFT band structure is implemented within the Wannier90 code \cite{wannier}. In TaAs, we construct the Hamiltonian in the Wannier basis using 20 d-orbitals of the 4 Ta atom and 12 p-orbitals of 4 As atoms to generate the localized Wannier functions. Overall, the Wannier-fitted band structure agrees well with the DFT one. And according to the above calculation results we select three sets of two-band of TaAs (with SOC) to calculate the inter-band nonlinear optical conductivity. One set of the results are shown in  Fig.~\ref{fig:TaAs} (c) and (d), the selected bands are marked with solid blue lines in  Fig.~\ref{fig:TaAs} (b) (the other two sets of the results see supplementary materials \cite{supp}). The calculated band structures of TaAs are symmetric, so their second order  intra-band nonlinear optical conductivity is zero. However with wannier fitting of the bands, we find the Hamiltonian (with SOC) is not symmetric. Based on this we calculate the inter-band nonlinear optical conductivity of TaAs. And it is a good example that intra-band nonlinear optical conductivity almost vanishes while inter-band nonlinear optical conductivity is nonzero.

\begin{table}[ht]
    \centering
    \caption{\label{Symm} The comparison of inter-band velocity matrix elements and  shift vector in the k and -k directions, the unit of velocity is $\mathrm{\mathring{A}}/\rm{s}.$} 
    \begin{ruledtabular}
    \begin{tabular}{c|c|c}
   
  $\rm{B_{8}}$ & k(1/14,1/14,0)  & -k(-1/14,-1/14,0) \tabularnewline  
    \hline 
    $v_{+-}$ & 0.07271 - 0.21689i & -0.07271 - 0.21689i \tabularnewline
    \hline
     $ R_{+ -}^{x, y}$& -0.04790 - 0.63258i & -0.04790 + 0.63258i \tabularnewline
    \hline
    TaAs & k(0,1/6,0) & -k(0,-1/6,0) \tabularnewline  
    \hline
    
    $v_{+-}$ & 1.16929e-05 & -1.16929e-05   \tabularnewline
    \hline
    
    $v_{+-}$(SOC) & 0.05099 + 0.02781i & -0.04359 - 0.02798i \tabularnewline
    \hline
    
    $\rm{LiNbO_{3}}$ & k(1/9,1/9,1/6) & -k(-1/9,-1/9,-1/6) \tabularnewline  
    \hline
    
    $v_{+-}$ & 0.08776 + 0.05603i & -0.08776 + 0.05603i \tabularnewline
    \hline
     $v_{+-}$(SOC) & 0.00066 + 0.00139i & 0.00301 + 0.00135i \tabularnewline

    \end{tabular}\end{ruledtabular}
\end{table}

In table II, we present the comparison of shift vector\cite{Rappe1,Rappe2,Fre,Jun} $R_{n m}^{x, y}=\frac{\partial \phi_{n m}^{y}}{\partial k_{x}}+\mathcal{L}_{n}^x-\mathcal{L}_{m}^x$ and the inter-band velocity matrix elements  in the k and -k directions. Here we use the expression  $R_{+ -}^{x, y}=-i\left[\frac{v_{+ -}^{x}  v^{y}_{\Delta}+v_{+ -}^{y}  v^{x}_{\Delta}}{v_{+ -}^{y} \varepsilon_{\Delta}}-  \frac{\tilde{m}_{+-}^{y x}}{v_{+ -}^{y}}\right]$ for two-band model to calculate the shift vector of $\rm{B_8}$, where  $v^{y}_{\Delta}=v_{+ +}^{y}-v_{- -}^{y}$, $\varepsilon_{\Delta}$ is the energy difference between the two bands and $\tilde{m}_{+-}^{x y}=\left\langle +\left|\partial_{k_{x}} \partial_{k_{y}} H\right| -\right\rangle$. The imaginary part of $\rm{B_8}$ shift vector is odd, so only the real part contributes to the solution of the third rank tensor of shift current response. More detailed information of velocity matrix elements and shift vector see supplementary materials \cite{supp}. A conduction band and a valence band closest to the Fermi surface are selected for calculation. 

\begin{figure}
    \centering
    \includegraphics[width=3.3in]{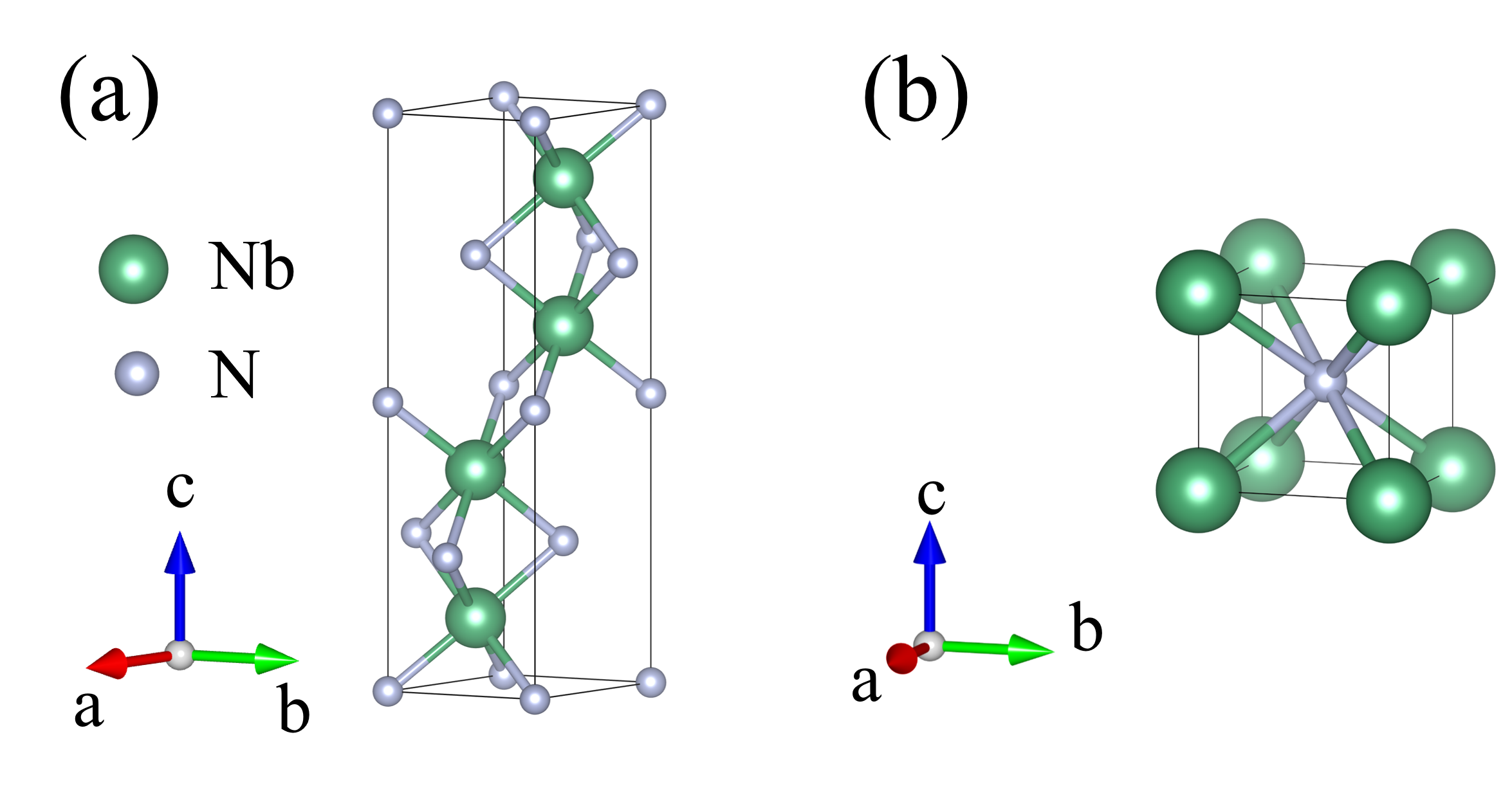}
    \caption{(a) Lattice structure of the hexagonal $\epsilon$-NbN ($\rm{P6_3}$/mmc, No. 194) with an estimate of $T_c$ = 10.230 K. (b) Lattice structure of the cubic NbN ($\rm{Pm\overline{3}m}$, No. 221) with an estimate of $T_c$ = 2.048 K.}
    \label{fig:FIG6}
\end{figure}

\begin{figure}
    \centering
    \includegraphics[width=3.3in]{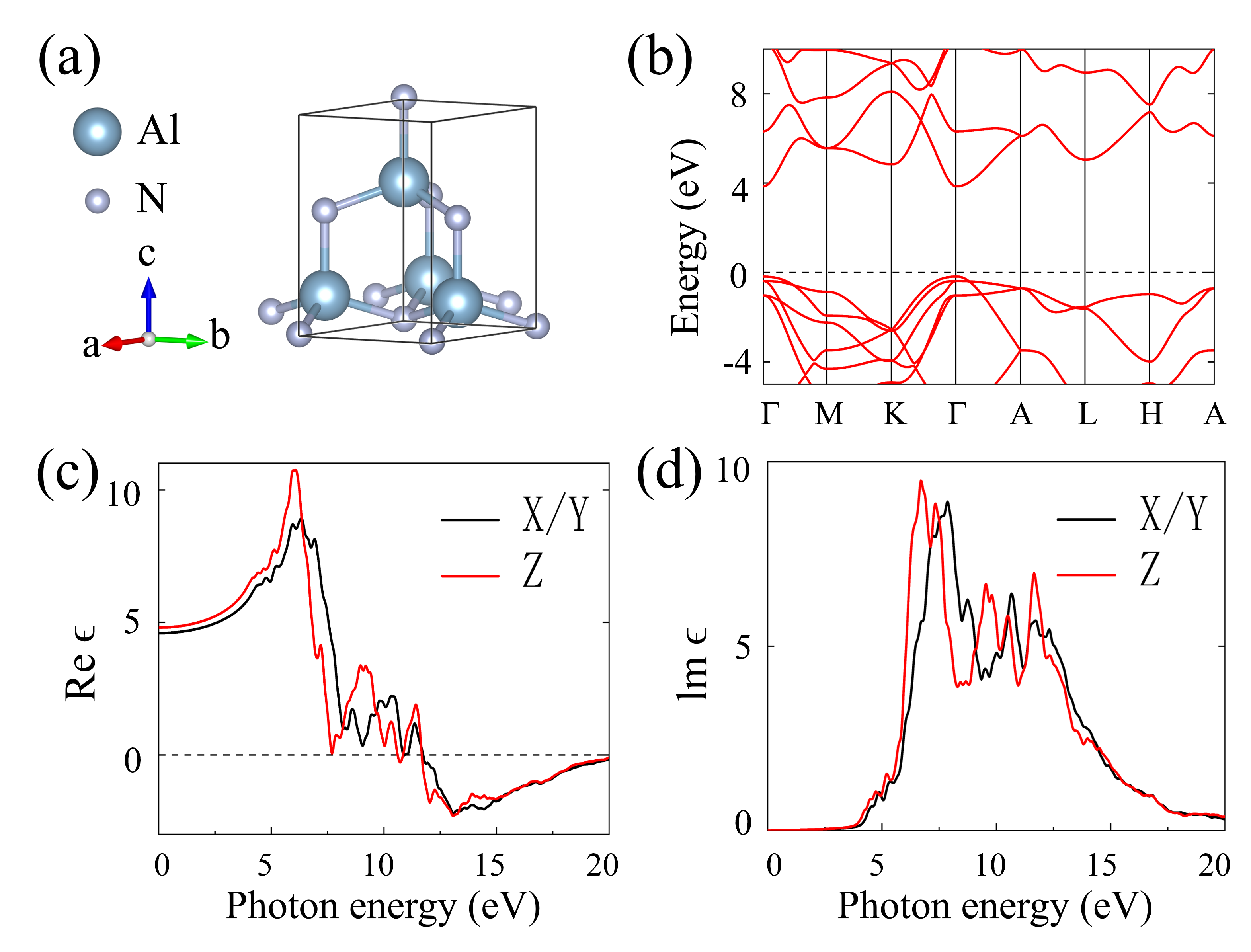}
    \caption{(a) The lattice structures of AlN, and the blue and grey atoms are Al and N. (b) The energy band of AlN. (c) The real part of the dielectric function of AlN along X,Y and Z direction. Slight anisotropy is observed. (d) The imaginary part of the dielectric function of AlN.}
    \label{fig:FIG7}
\end{figure}

In a Josephson junction, the inter-layer supercurrent  $J_s$ is determined from the phase difference $\phi$ of the two superconducting layers, $J_s=J_c\rm{sin}(\phi)$, quite different from the conventional current $J=\sigma E$. Electric and magnetic field in layered superconductors (Josephson junctions) form plane-wave-like Josephson plasma waves (JPW) or soliton-like Josephson vortices (JV), both are solutions of the sine-Gordon equations of $\phi$ \cite{Savel}. The Josephson plasma frequency $\omega_J$ is in the Terahertz range. Because of $\rm{sin}(\phi)\approx\phi-\phi^3/6$, nonlinear JPW could propagate below, and very close to $\omega_J$. Coherent emission from JPW \cite{JSPTHZ} or Cherenkov radiation from JV lattices are proposed. The current-current-current correlation is known as the third moment of shot noise of the Josephson junction, and changes the average thermal escape rate \cite{I3}. The Josephson plasma frequency $\omega_{J}=\sqrt{\frac{8 \pi n e D J_{c}}{\hbar \epsilon}}$ \cite{Savel}, where $D$  is the total thickness of an insulating layer and a superconducting layer,  $\epsilon$ is the dielectric constant of the insulating layers, $n=N/2$, $N$ is the electron number of a superconducting quasiparticle, for Cooper pair, $N=2$. $J_c$ is the critical current density of the junction structure. 

In Fig.~\ref{fig:FIG6}, we present the structure of (a) hexagonal $\varepsilon$-NbN and (b) the cubic NbN \cite{Kim,Espiau}. The superconducting transition temperature ($T_c$) is determined by Allen-Dynes modified McMillan equation \cite{Elia,Allen,Maym,tc},
$k_B T_c=\frac{\hbar\omega_{\rm{log}}}{1.2}\mathrm{exp}\left[\frac{-1.04(1+\lambda)}{\lambda(1-0.62\mu^*)-\mu^*}\right]$, with the parameters estimated from Quantum ESPRESSO.
For hexagonal $\epsilon$-NbN the logarithmic average $\omega_{\rm{log}}$ = 350.492 K (30.244 meV), the electron-phonon coupling $\lambda$ = 0.65364 and the Coulomb repulsion $\mu^{*}$ = 0.10,
for cubic-NbN, $\omega_{\rm{log}}$ = 623.388K (53.792 meV), $\lambda$ = 0.41705 and $\mu^{*}$ = 0.12. The pressure is at 0.1 MPa. It is found that $T_c$ = 10.230 K for hexagonal $\epsilon$-NbN and $T_c$ = 2.048 K for cubic-NbN. In Fig.~\ref{fig:FIG7}, we present the structure, band and the dielectric function of insulating layer $\rm{AlN}$. The dielectric function along the x/y axis is different from that along the z axis. The real part of the dielectric function becomes negative in the high frequency region.

\begin{figure}
    \centering
    \includegraphics[width=3.3in]{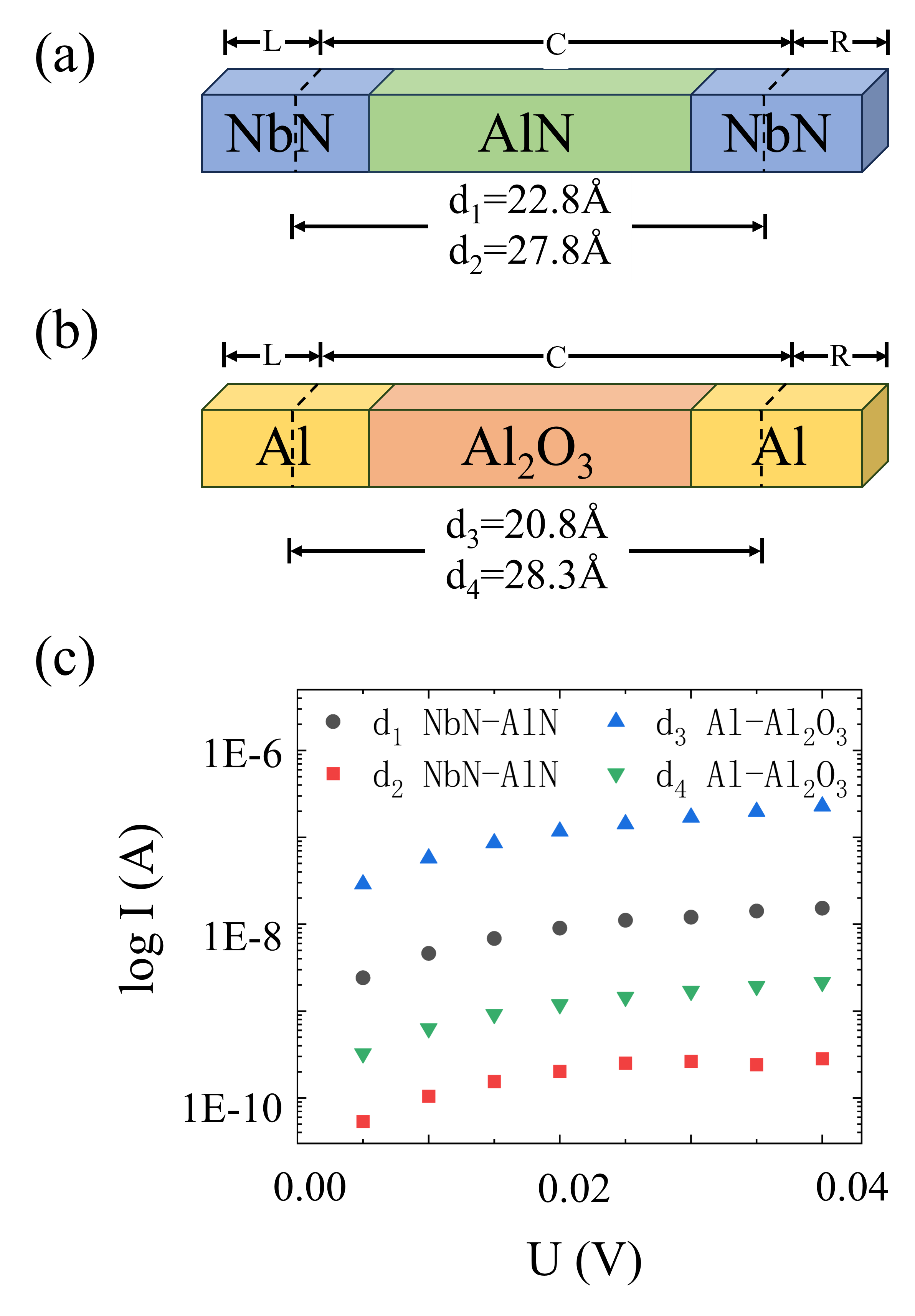}
    \caption{Schematic of (a) NbN-AlN-NbN and (b) Al-Al$_2$O$_3$-Al heterojunction. (c) The I-V curve of the four devices with different length of the central region. The resistance calculated from the I-V curve is important for optimizing the quality factor of the Josephson junction cavity. As the heat dissipated from the cavity is directly controlled by the resistance. }
    \label{fig:FIG8}
\end{figure}

In Fig.~\ref{fig:FIG8}, We provide schematic of the NbN-AlN-NbN and Al-Al$_2$O$_3$-Al hetero-junction devices. The details of the structure is provided in the supplementary material. The device is divided into three parts: the left electrode (L), the central scattering region (C), and the right electrode (R). In Fig.~\ref{fig:FIG8} (c), we observe that for similar thickness of the central scattering region, the resistance of NbN-AlN-NbN junction is significantly larger than that of the Al-Al$_2$O$_3$-Al junction, as one compares the black dots and the blue triangle. As the thickness increases, the resistance of both devices increases. For a given voltage, larger resistance means smaller dissipation of heat.

\begin{figure}
    \centering
    \includegraphics[width=3.2in]{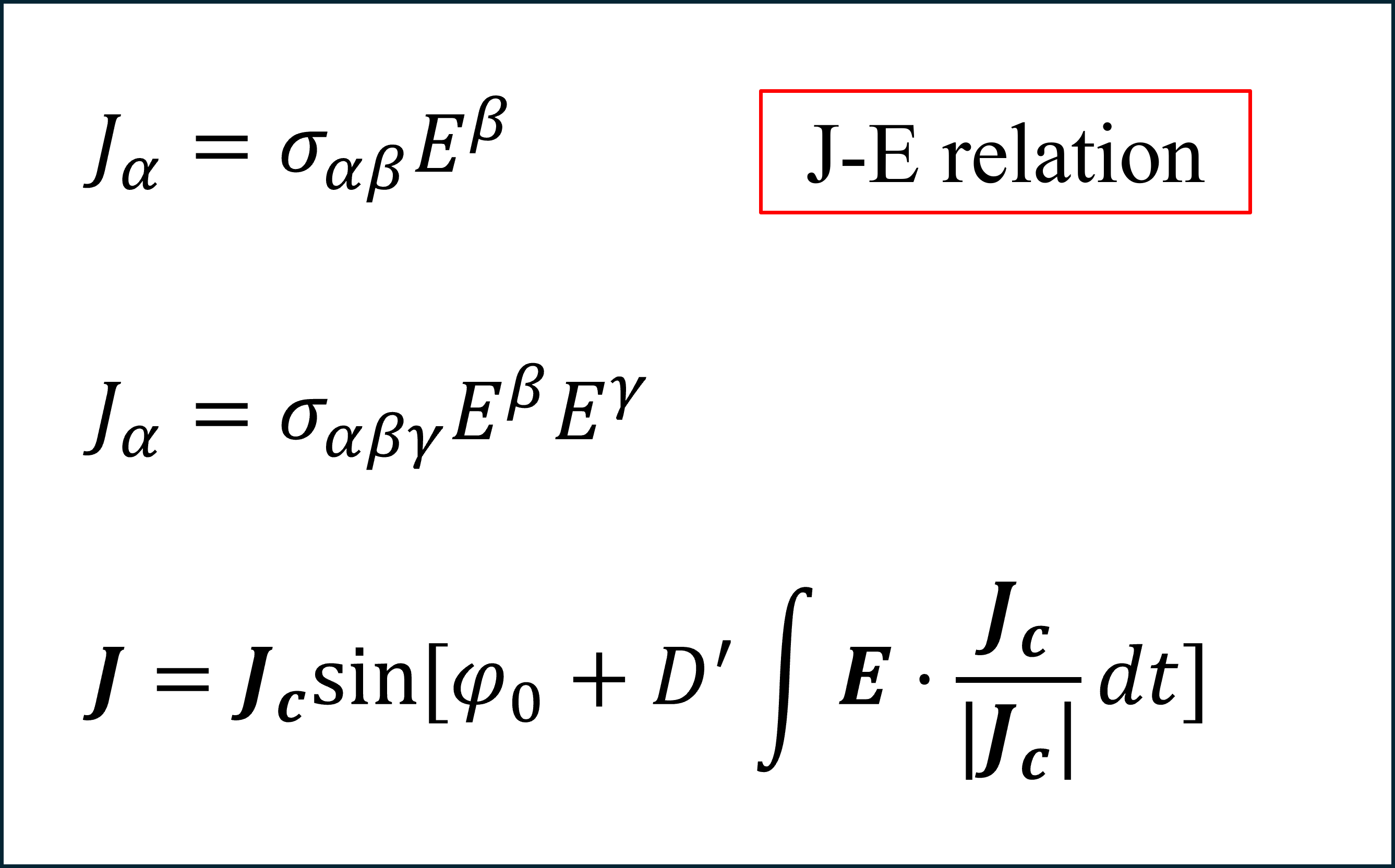}
    \caption{The relationship between the electric current and the electric field.} 
    \label{fig:relation}
\end{figure}

To conclude, we develop the velocity-operator approach to calculate the nonlinear optical conductivity from a current-current-current correlation. We discuss the relationship between the electric current and the electric field in several different cases (see Fig.~\ref{fig:relation}): linear, nonlinear, and superconductor Josephson junction. In the formula for superconductor Josephson junction, $D^{\prime}$ is equal to 2eD/$\hbar$, so the unit of $D^{\prime}$ is m/V·s, which makes $D^{\prime} \int E \cdot \frac{\boldsymbol{J}_{c}}{\left|\boldsymbol{J}_{\boldsymbol{c}}\right|} d t$ a dimensionless phase. Superconductor Josephson junction has a unique nonlinear “current-electric field” relation, which is used to design Terahertz laser sources. And the nonlinear optical conductivity is also used to design Terahertz laser sources. We establish connections between our approach and the position-operator approach. We investigate both the intra-band and inter-band contributions to the nonlinear optical conductivity, which are computable from a wannier Hamiltonian fitted from DFT. The intra-band is not emphasized in the visible and ultraviolet frequency band, but has a strong sum-frequency generation in the THz frequency band, predicted by a titled Dirac cone model. The abrupt change of velocity and the asymmetry in the momentum space are the key requirement for such an intra-band process. Similar mechanism is used in the step recovery diode to do a up-conversion of frequency from GHz to THz. Nonlinear Josephson plasmon devices also oscillate in the THz frequency region. The superconductor transition temperature and the dielectric function of the insulator are analyzed for these devices. The resistance of the NbN-AlN-NbN and Al-Al$_2$O$_3$-Al hetero-junction devices are calculated.
\begin{acknowledgments}
The authors thank Prof. Hu Xu for helpful discussions on the density functional theory, and Ms. Yang Liu for providing an estimate of the superconducting transition temperature of two different structures of NbN. This work is supported by National Natural Science Foundation of China (61988102), the Key Research and Development Program of Guangdong Province (2019B090917007) and the Science and Technology Planning Project of Guangdong Province (2019B090909011). Z. L. acknowledges the support of funding from Chinese Academy of Science E1Z1D10200 and E2Z2D10200; from ZJ project 2021QN02X159 and from JSPS Grant No. PE14052
and P16027.\end{acknowledgments}


\begin{thebibliography}{10}

\bibitem{Heinz1} A. Nahata, A.S. Weling and T.F. Heinz, A wideband coherent terahertz spectroscopy system using optical rectification and electro-optic sampling, Appl. Phys. Lett. \textbf{69}, 2321-2323 (1996).

\bibitem{Zhangxc} Q. Wu and X. C. Zhang, Free-space electro-optic sampling of mid-infrared pulses, Appl. Phys. Lett. \textbf{71}, 1285-1286 (1997).

\bibitem{Heb}J\'anos Hebling, G\'abor Alm\'asi, Ida Z. Kozma, and J\"urgen Kuhl, Velocity matching by pulse front tilting for large-area THz-pulse generation, Optics Express \textbf{10} (21), 1161-1166 (2002).

\bibitem{Rappe1}S. M. Young and A. M. Rappe, First Principles Calculation
of the Shift Current Photovoltaic Effect in Ferroelectrics, Phys.
Rev. Lett. \textbf{109}, 116601 (2012).

\bibitem{Rappe2}S. M. Young, F. Zheng and A. M. Rappe, First-Principles
Calculation of the Bulk Photovoltaic Effect in Bismuth Ferrite, Phys. Rev. Lett. \textbf{109}, 236601 (2012).


\bibitem{Grinberg}I. Grinberg, D. V. West, M. Torres, G. Gou, D.
M. Stein, L. Wu, G. Chen, E. M. Gallo, A. R. Akbashev, P. K. Davies,
J. E. Spanier, A. M. Rappe, Perovskite oxides for visible light-absorbing
ferroelectric and photovoltaic materials, Nature \textbf{503}, 509\textendash 512
(2013).

\bibitem{Nie}W. Nie, H. Tsai, R. Asadpour, J.-C. Blancon, A. J. Neukirch,
G. Gupta, J. J. Crochet, M. Chhowalla, S. Tretiak, M. A. Alam, H.-L.
Wang, A. D. Mohite, High-efficiency solution processed perovskite
solar cells with millimeter-scale grains, Science \textbf{347}, 522\textendash 525
(2015).

\bibitem{Shi}D. Shi, V. Adinolfi, R. Comin, M. Yuan, E. Alarousu,
A. Buin, Y. Chen, S. Hoogland, A. Rothenberger, K. Katsiev, Y. Losovyj,
X. Zhang, P. A. Dowben, O. F. Mohammed, E. H. Sargent, O. M. Bakr,
Low trap-state density and long carrier diffusion in organolead trihalide perovskite single crystals, Science \textbf{347}, 519\textendash 522
(2015).

\bibitem{Quile}D. W. de Quilettes, S. M. Vorpahl, S. D. Stranks,
H. Nagaoka, G. E. Eperon, M. E. Ziffer, H. J. Snaith, D. S. Ginger,
Impact of microstructure on local carrier lifetime in perovskite solar
cells, Science \textbf{348}, 683\textendash 686 (2015).
\bibitem{Rappe3}L. Z. Tan and A. M. Rappe, Enhancement of the Bulk
Photovoltaic Effect in Topological Insulators, Phys. Rev. Lett. \textbf{116},
237402 (2016).

\bibitem{Cook} A. M. Cook, B. M. Fregoso, F. de Juan, S. Coh and
J. E. Moore, Design principles for shift current photovoltaics, Nature
Communications \textbf{8}, 14176 (2017).

\bibitem{Franken}P. A. Franken, A. E. Hill, C. W. Peters, and G.
Weinreich, Generation of Optical Harmonics, Phys. Rev. Lett. \textbf{7},
118 (1961).

\bibitem{Heinz}H. W. K. Tom, T. F. Heinz, and Y. R. Shen, Second-Harmonic
Reflection from Silicon Surfaces and Its Relation to Structural Symmetry,
Phys. Rev. Lett. \textbf{51}, 1983 (1983).

\bibitem{Savel} S. Savel'ev, A. L. Rakhmanov, V. A. Yampol'skii,
F. Nori, Analogues of nonlinear optics using terahertz Josephson plasma
waves in layered superconductors, Nature Physics \textbf{2}, 521-525
(2006)

\bibitem{Kockum} A. F. Kockum, A. Miranowicz, V. Macr\'{i}, S. Savasta,
F. Nori, Deterministic quantum nonlinear optics with single atoms
and virtual photons, Phys. Rev. A \textbf{95}, 063849 (2017).

\bibitem{Kockum1} A. F. Kockum, V. Macr\'{i}, L. Garziano, S. Savasta,
F. Nori, Frequency conversion in ultrastrong cavity QED, Scientific
Reports \textbf{7}, 5313 (2017).

\bibitem{Stassi} R. Stassi, V. Macr\'{i}, A. F. Kockum, O.D. Stefano,
A. Miranowicz, S. Savasta, F. Nori, Quantum Nonlinear Optics without
Photons, Phys. Rev. A \textbf{96}, 023818 (2017).

\bibitem{Gu}X. Gu, A. F. Kockum, A. Miranowicz, Y.X. Liu, F. Nori,
Microwave photonics with superconducting quantum circuits Physics
Reports 718-719, pp. 1-102 (2017).

\bibitem{Sipe}J. E. Sipe and E. Ghahramani, Nonlinear optical response
of semiconductors in the independent-particle approximation, Phys.
Rev. B \textbf{48}, 11705 (1993).

\bibitem{Mikhailov}S. A. Mikhailov, Theory of the giant plasmon-enhanced
second-harmonic generation in graphene and semiconductor two-dimensional
electron systems, Phys. Rev. B. \textbf{8}4, 045432 (2011).

\bibitem{Daria}Daria Smirnova and Yuri S. Kivshar, Second-harmonic
generation in subwavelength graphene waveguides, Phys. Rev. B. \textbf{90},
165433 (2014).

\bibitem{Bloemb} N. Bloembergen, Nonlinear optics (W. A. Benjamin
Inc., New York, 1965).

\bibitem{Fre} B. M. Fregoso, T. Morimoto, and J. E. Moorel, Quantitative relationship between polarization differences and the zone-averaged shift photocurrent, Phys. Rev. B \textbf{96}, 075421 (2017).

\bibitem{Jun} J. Ahn, G.-Y. Guo and N. Nagaosa, Low-Frequency Divergence and Quantum Geometry of the Bulk Photovoltaic Effect in Topological Semimetals, Phys. Rev. X \textbf{10}, 041041 (2020).

\bibitem{Supp} See Supplemental Material at {[}URL{]} for the discussion
of the general formula of the nonlinear conductivity tensor, the expansion
of the imaginary frequency Green's function into a sum of the real
frequency spectral function, the calculation of the sum in imaginary
frequency, and the evaluation of the trace of the matrix elements.

\bibitem{Nori_Li}Zhou Li and F. Nori, Nonlinear response in a noncentrosymmetric topological insulator, Phys. Rev. B \textbf{99}, 155146 (2019).

\bibitem{VPGusynin} V. P. Gusynin, S. G. Sharapov, J. P. Carbotte, Sum rules for the optical and Hall conductivity in graphene, Phys. Rev. B \textbf{75}, 165407 (2007).

\bibitem{LiZhou} Zhou Li and J. P. Carbotte, Optical spectral weight: Comparison of weak and strong spin-orbit coupling, Phys. Rev. B \textbf{91}, 115421 (2015).

\bibitem{Sipe1}Claudio Aversa and J. E. Sipe, Nonlinear optical susceptibilities of semiconductors: Results with a length-gauge analysis, Phys. Rev. B \textbf{52}, 14636 (1995).

\bibitem{Jiang} X. Jiang, L. Kang, J. Wang, B. Huang, Giant Bulk Electrophotovoltaic Effect in Heteronodal-Line Systems, Phys. Rev. Lett. \textbf{130} (25), 256902 (2023).

\bibitem{Li22} X. Huang, X. Jiang, B. Huang,and Z. Li, Nonlocal optical conductivity of Fermi surface nesting materials, Sci. China Phys. Mech. \textbf{63}, 1 (2022).

\bibitem{Fu} L. Fu, Hexagonal Warping Effects in the Surface States
of the Topological Insulator $\textrm{Bi}_{2}\textrm{Te}_{3}$, Phys.
Rev. Lett \textbf{103}, 266801 (2009).

\bibitem{Li2}Zhou Li and J. P. Carbotte, Hexagonal warping on spin
texture, Hall conductivity, and circular dichroism of topological
insulators, Phys. Rev. B \textbf{89}, 165420 (2014).

\bibitem{Li1} Zhou Li and J. P. Carbotte, Hexagonal warping on optical
conductivity of surface states in topological insulator $\textrm{Bi}_{2}\textrm{Te}_{3}$,
Phys. Rev. B \textbf{87}, 155416 (2013).

\bibitem{Xu1} S. Y. Xu et al., Topological Phase Transition and Texture
Inversion in a Tunable Topological Insulator, Science \textbf{332},
560 (2011).

\bibitem{Xu2} S. Y. Xu et al., Hedgehog spin texture and Berry's
phase tuning in a magnetic topological insulator, Nature Phys. \textbf{8},616
(2012).

\bibitem{Chen1} Y. L. Chen et.al, Massive Dirac Fermion on the Surface
of a Magnetically Doped Topological Insulator, Science \textbf{329},
659 (2010).

\bibitem{Hasan} M. Z. Hasan and C. L. Kane, Colloquium: Topological
insulators, Rev. Mod. Phys. \textbf{82}, 3045 (2010).

\bibitem{Qi1} X.-L. Qi and S.-C. Zhang, Topological insulators and
superconductors, Rev. Mod. Phys. \textbf{83}, 1057 (2011).

\bibitem{Moore} J. E. Moore, The birth of topological insulators,
Nature \textbf{464}, 194 (2010).

\bibitem{Hsieh1} D. Hsieh et al., Observation of Unconventional Quantum
Spin Textures in Topological Insulators, Science \textbf{323}, 919
(2009).

\bibitem{Chen} Y. L. Chen et al., Experimental Realization of a Three-Dimensional
Topological Insulator, $\textrm{Bi}_{2}\textrm{Te}_{3}$, Science
\textbf{325}, 178 (2009).

\bibitem{Hsieh2} D. Hsieh et al., A tunable topological insulator
in the spin helical Dirac transport regime, Nature(London) \textbf{460},
1101 (2009).

\bibitem{Tokura}K. N Okada, Y. Takahashi, M. Mogi, R. Yoshimi, A.
Tsukazaki, K. S. Takahashi, N. Ogawa, M. Kawasaki and Y. Tokura, Observation
of topological Faraday and Kerr rotations in quantum anomalous Hall
state by terahertz magneto-optics. Nat. Commun. \textbf{7}, 12245 (2016).

\bibitem{Yasuda}K. Yasuda, A. Tsukazaki, R. Yoshimi, K. S. Takahashi,
M. Kawasaki, and Y. Tokura, Large Unidirectional Magnetoresistance
in a Magnetic Topological Insulator. Phys. Rev. Lett. \textbf{117},
127202 (2016).


\bibitem{Lizy} Zi-Yuan Li, Qi Li and Zhou Li, High-order harmonic generations in tilted Weyl semimetals, Chinese Phys. B \textbf{31} 124204 (2022).


\bibitem{Kresse} G. Kresse, D. Joubert, From ultrasoft pseudopotentials to the projector augmented wave method
 Phys. Rev. B \textbf{59}, 1758-1775 (1999).

\bibitem{Blochl} P. E. Blochl, Phys. Rev. B \textbf{50}, 17953-17979 (1994).

\bibitem{Perdew} J. P. Perdew, K. Burke, M. Ernzerhof, Generalized Gradient Approximation Made Simple, Phys. Rev. Lett. \textbf{77}, 3865-3868 (1996).

\bibitem{Grimme} S. Grimme, Semiempirical GGA-type density functional constructed with a long-range dispersion correction, J. Comput. Chem. \textbf{27}, 1787-1799 (2006).

\bibitem{sg1} K. Lejaeghere, G. Bihlmayer, T. Björkman, et al, Reproducibility in density functional theory calculations of solids, Science, \textbf{351}(6280) (2016).

\bibitem{sg2} D. R. Hamann, Optimized norm-conserving Vanderbilt pseudopotentials, Physical Review B, \textbf{88}(8), 085117 (2013).

\bibitem{Bellaiche} L. Bellaiche and D. Vanderbilt, Virtual crystal approximation revisited: Application to dielectric and piezoelectric properties of perovskites, Phys. Rev. B \textbf{61}, 7877 (2000).

\bibitem{Lei} B. H. Lei, S. Pan, Z. Yang, C. Cao, D. J. Singh, Second harmonic generation susceptibilities from symmetry adapted Wannier functions, Phys. Rev. Lett. \textbf{125}, 187402 (2020).

\bibitem{vasp1} G. Kresse and J. Hafner, Ab initio molecular dynamics for openshell transition metals, Phys. Rev. B \textbf{48}, 13115 (1993).

\bibitem{vasp2} G. Kresse and J. Furthmüller, Efficiency of ab-initio total energy calculations for metals and semiconductors using a plane-wave basis set. Comp. Mat. Sci. \textbf{6}, 15–50 (1996).

\bibitem{wannier} A. A.  Mostofi, J. R. Yates, G. Pizzi, Y. S. Lee, I. Souza, D. Vanderbilt, N. Marzari, An updated version of wannier90: A tool for obtaining maximally-localised Wannier functions, Comput. Phys. Commun. \textbf{185}, 2309 (2014).

\bibitem{supp} See supplementary materials for: "Design principles of nonlinear optical materials for Terahertz lasers” for details of the inter-band nonlinear optical conductivity and shift vector.

\bibitem{JSPTHZ} D. Nicoletti, M. Buzzi, M. Fechner, P. E. Dolgirev, M. H. Michael, J. B. Curtis, E. Demler, G. D. Gu, and A. Cavalleri, Coherent emission from surface Josephson plasmons in striped cuprates, PNAS 119, 39, e2211670119 (2022).

\bibitem{I3} A. V. Timofeev, M. Meschke, J. T.Peltonen, T. T. Heikkila and J. P. Pekola, Wideband Detection of the Third Moment of Shot Noise by a Hysteretic Josephson Junction, Phys. Rev. Lett. \textbf{98}, 207001 (2007).

\bibitem{Kim} S. Kim, H. Terai, T. Yamashita, et al. Enhanced coherence of all-nitride superconducting qubits epitaxially grown on silicon substrate, Communications Materials 
 \textbf{2}, 98 (2021).

\bibitem{Espiau} R. Espiau de Lamaëstre, P. Odier, J. C. Villégier, Microstructure of NbN epitaxial ultrathin films grown on A-, M-, and R-plane sapphire, Appl. Phys. Lett. \textbf{91}, 232501 (2007).

\bibitem{Elia} G. M. Eliashberg, Interactions between electrons and lattice vibrations in a superconductor, Sov. Phys. JETP \textbf{11} (3), 696-702 (1960).

\bibitem{Allen} P. B. Allen, Neutron spectroscopy of superconductors, Phys. Rev. B \textbf{6}(7), 2577 (1972).

\bibitem{Maym} Yansun Yao, J. S. Tse, K. Tanaka, F. Marsiglio, and Y. Ma, Superconductivity in lithium under high pressure investigated with density functional and Eliashberg theory, Phys. Rev. B \textbf{79}, 054524 (2009).

\bibitem{tc} F. Giustino, Electron-phonon interactions from first principles, Rev. Mod. Phys. \textbf{89}(1), 015003 (2017).

 

\end{thebibliography}
\end{document}